\date{}
\documentclass[11pt]{article}
\usepackage{amsmath,amsfonts,latexsym,graphicx,amssymb}
\usepackage{amsfonts}
\usepackage{bm}

\newcommand{\ot}{{\,\otimes\,}}
\newcommand{{\Cd}}{{\mathbb{C}^d}}

\def\oper{{\mathchoice{\rm 1\mskip-4mu l}{\rm 1\mskip-4mu l}%
{\rm 1\mskip-4.5mu l}{\rm 1\mskip-5mu l}}}
\def\<{\langle}
\def\>{\rangle}
\newtheorem{theorem}{Theorem}

\begin{document}

\title{\textbf{On the structure of entanglement witnesses \\
and new class of positive  indecomposable maps}}
\author{Dariusz
Chru\'sci\'nski\thanks{email: darch@phys.uni.torun.pl}$\ $ and
Andrzej
Kossakowski \\
Institute of Physics, Nicolaus Copernicus University,\\
Grudzi\c{a}dzka 5/7, 87--100 Toru\'n, Poland}

\maketitle

\begin{abstract}

We  construct a new class of positive indecomposable maps in the
algebra of $d \times d$ complex matrices.  Each map is uniquely
characterized by a cyclic bistochastic matrix. This class
generalizes a Choi map for $d=3$. It provides a new reach family
of indecomposable entanglement witnesses which define important
tool for investigating quantum entanglement.

\end{abstract}


\section{Introduction}

One of the most important problems of quantum information theory
\cite{QIT} is the characterization of mixed states of composed
quantum systems. In particular it is of primary importance to test
whether a given quantum state exhibits quantum correlation, i.e.
whether it is separable or entangled. For low dimensional systems
there exists simple necessary and sufficient condition for
separability. The celebrated Peres-Horodecki criterium
\cite{Peres,PPT} states that a state of a bipartite system living
in $\mathbb{C}^2 \ot \mathbb{C}^2$ or $\mathbb{C}^2 \ot
\mathbb{C}^3$ is separable iff its partial transpose is positive.
Unfortunately, for higher-dimensional systems there is no single
{\it universal} separability condition. A different useful
separability criterion, that has been used to show entanglement of
PPT states, is the range criterion \cite{PPT}. It is based on the
fact that for every separable state $\rho$ there exist a set of
pure product states $\psi_i \ot \varphi_i$ that span the range of
$\rho$ while $\psi_i \ot \overline{\varphi}_i$ span the range of
its partial transposition $(\oper \ot \tau)\rho$. Other criteria,
that are in general weaker than PPT are the reduction criterion
\cite{C1} and the majorization criterion \cite{C2}. None of these
criteria, nor a combination of them are sufficient to give a
complete characterization of separable states.

The most general approach to separability problem is based on the
following theorem \cite{Horodeccy-PM}: a state $\rho$ of a
bipartite system living in $\mathcal{H}_A \ot \mathcal{H}_B$ is
separable iff $\mbox{Tr}(W\rho) \geq 0$ for any Hermitian operator
$W$ satisfying $\mbox{Tr}(W P_A \ot P_B)\geq 0$, where $P_A$ and
$P_B$ are projectors acting on $\mathcal{H}_A$ and
$\mathcal{H}_B$, respectively.
 Recall, that a Hermitian  operator $W \in
\mathcal{B}(\mathcal{H}_A \ot \mathcal{H}_B)$ is an entanglement
witness \cite{Horodeccy-PM,Terhal1} iff: i) it is not positively
defined, i.e. $W \ngeq 0$, and ii) $\mbox{Tr}(W\sigma) \geq 0$ for
all separable states $\sigma$. A bipartite state $\rho$ living in
$\mathcal{H}_A \ot \mathcal{H}_B$ is entangled iff  there exists
an entanglement witness $W$ detecting $\rho$, i.e. such that
$\mbox{Tr}(W\rho)<0$. It should be stressed that there is no
universal $W$, i.e. there is no entanglement witness which detects
all entangled states. Each entangled state $\rho$ may be detected
by a specific choice of $W$.  It is clear that each $W$ provides a
new separability test and it may be interpreted as a new type of
Bell inequality \cite{W-Bell}. There is, however, no general
procedure for constructing  $W$'s.

The separability problem may be equivalently formulated in terms
positive maps \cite{Horodeccy-PM}: a state $\rho$ is separable iff
$(\oper \ot \Lambda)\rho$ is positive for any positive map
$\Lambda$ which sends positive operators on $\mathcal{H}_B$ into
positive operators on $\mathcal{H}_A$.  Unfortunately, in spite of
the considerable effort, the structure of positive maps is rather
poorly understood [9--44].

Note, that performing a PPT test we may reduce the separability
problem to PPT states. Positive maps which may be used to detect
PPT entangled states define a class of so called indecomposable
positive maps. There are only few examples of indecomposable
positive maps known in the literature (see review in Section
\ref{REVIEW}). They seem to be hard to find and no general
construction method is available. Therefore, any new example
provides new tools to investigate quantum entanglement.  In the
present paper we construct a new class of such maps. The paper is
organized as follows: in the next Section we introduce a natural
hierarchy of positive convex cones in the space of (unnormalized)
states of bipartite $d \ot d$ quantum systems. In Section
\ref{MAPS} we recall basis notions from the theory of positive
maps and introduce a duality between positive maps entanglement
witnesses. Section \ref{REVIEW} serves as a catalog of known
indecomposable positive maps. Sections \ref{CLASS} and
\ref{CONTRACTIONS} introduce basic classes of positive maps which
we are going to use in our search for indecomposable maps. Finally
in Section \ref{MAIN} we show how to construct  a new family of
indecomposable maps within a class of positive maps discussed in
previous sections. This class defines a natural generalization of
Choi maps on $M_3$. A brief discussion is included in the last
section.

\section{The structure of entanglement witnesses}

In the preset paper we shall consider a bipartite quantum system
living in $\Cd \ot \Cd$. Denote by $M_d$ a set of $d \times d$
complex matrices and let $M_d^+$ be a convex set of semi-positive
elements in $M_d$, that is, $M_d^+$ defines a space of
(unnormalized) states of $d$-level quantum system. For any $\rho
\in (M_d \ot M_d)^+$ denote by  $\mathrm{SN}(\rho)$  a Schmidt
number of $\rho$ \cite{SN}. Now, let us introduce the following
family of  positive cones:
\begin{equation}\label{}
    \mathrm{V}_r = \{\, \rho \in (M_d \ot M_d)^+\ |\
    \mathrm{SN}(\rho) \leq r\, \}  \ .
\end{equation}
One has the following chain of inclusions
\begin{equation}\label{V-k}
\mathrm{V}_1  \subset \ldots \subset  \mathrm{V}_d \equiv (M_d \ot
M_d)^+\ .
\end{equation}
Clearly, $\mathrm{V}_1$ is a cone of separable (unnormalized)
states and $V_d \smallsetminus  V_1$ stands for a set of entangled
states. Note, that a partial transposition $(\oper_d \ot \tau)$
gives rise to another family of cones:
\begin{equation}\label{}
     \mathrm{V}^l = (\oper_d \ot \tau)\mathrm{V}_l \ ,
\end{equation}
such that $ \mathrm{V}^1  \subset \ldots \subset  \mathrm{V}^d$.
 One has
$\mathrm{V}_1 = \mathrm{V}^1$, together with the following
hierarchy of inclusions:
\begin{equation}\label{}
    \mathrm{V}_1 = \mathrm{V}_1 \cap \mathrm{V}^1 \subset \mathrm{V}_2 \cap
    \mathrm{V}^2   \subset \ldots \subset \mathrm{V}_d \cap \mathrm{V}^d \
    .
\end{equation}
Note, that $\mathrm{V}_d \cap \mathrm{V}^d$ is a convex set of PPT
(unnormalized) states. Finally, $\mathrm{V}_r \cap \mathrm{V}^s$
is a convex subset of PPT states $\rho$  such that
$\mathrm{SN}(\rho) \leq r$ and $\mathrm{SN}[(\oper_d \ot
\tau)\rho] \leq s$.

Let us denote by $\mathrm{W}$ a space of entanglement witnesses,
i.e. a space of non-positive Hermitian operators $W \in M_d \ot
M_d$ such that  $\mathrm{Tr}(W\rho)\geq 0$ for all $\rho \in
\mathrm{V}_1$.  Define a family of subsets $\mathrm{W}_r \subset
M_d \ot M_d$:
\begin{equation}\label{}
\mathrm{W}_r = \{\, W\in M_d \ot M_d\ |\ \mathrm{Tr}(W\rho) \geq
0\ , \ \rho \in \mathrm{V}_r\, \}\ .
\end{equation}
One has
\begin{equation}\label{}
(M_d \ot M_d)^+ \equiv \mathrm{W}_d  \subset \ldots \subset
\mathrm{W}_1   \ .
\end{equation}
Clearly, $\mathrm{W} = \mathrm{W}_1 \smallsetminus \mathrm{W}_d$.
Moreover, for any $k>l$, entanglement witnesses from $\mathrm{W}_l
\smallsetminus \mathrm{W}_k$  can detect entangled states from
$\mathrm{V}_k \smallsetminus  V_l$, i.e. states $\rho$ with
Schmidt number $l < \mathrm{SN}(\rho) \leq k$. In particular $W
\in \mathrm{W}_k \smallsetminus \mathrm{W}_{k+1}$ can detect state
$\rho$ with $\mathrm{SN}(\rho)=k$.

Consider now the following class
\begin{equation}\label{}
    \mathrm{W}_r^s = \mathrm{W}_r + (\oper \ot \tau)\mathrm{W}_s\
    ,
\end{equation}
that is, $W \in \mathrm{W}_r^s$ iff
\begin{equation}\label{}
    W = P + (\oper \ot \tau)Q\ ,
\end{equation}
with $P \in \mathrm{W}_r$ and $Q \in \mathrm{W}_s$. Note, that
$\mathrm{Tr}(W\rho) \geq 0$ for all $\rho \in \mathrm{V}_r \cap
\mathrm{V}^s$. Hence such $W$ can detect PPT states $\rho$ such
that $\mathrm{SN}(\rho) \geq r$ or $\mathrm{SN}[(\oper_d \ot
\tau)\rho] \geq s$. Entanglement witnesses from $\mathrm{W}_d^d$
are called decomposable \cite{optimal}. They cannot detect PPT
states. One has the following chain of inclusions:
\begin{equation}\label{}
    \mathrm{W}_d^d \subset \ldots \subset \mathrm{W}^2_2 \subset \mathrm{W}^1_1 \equiv
    \mathrm{W}\ .
\end{equation}
The `weakest' entanglement can be detected by elements from
$\mathrm{W}_1^1 \smallsetminus \mathrm{W}_2^2$. We shall call them
{\em atomic entanglement witnesses}.

\section{Positive maps and duality}
\label{MAPS}

It is well known that the separability problem may be reformulated
in terms of positive maps \cite{Horodeccy-PM}. Recall, that
 a linear map
$\varphi : M_d \longrightarrow M_d$ is called positive  iff
$\varphi(a) \in M_d^+$ for any $a \in M_d^+$. It is well known
\cite{Horodeccy-PM} that a state $\rho \in (M_d \ot M_d)^+$ is
separable iff $(\oper_d \ot \varphi)\rho \geq 0$ for all positive
maps $\varphi : M_d \longrightarrow M_d$ ($\oper_d$ stands for an
identity map). Hence, having a positive map $\varphi$ such that
$(\oper_d \ot \varphi)$ acting on $\rho$ is no longer positive we
are sure that $\rho$ is entangled. However, the crucial problem
with the above criterion is that the classification and
characterization of positive maps is an open question.

A linear map $\varphi : M_d \longrightarrow M_d$ is called
$k$-positive iff the extended map
\[ \oper_k \ot \varphi : M_k \ot
M_d \longrightarrow M_k \ot M_d\ , \] is positive. If $\varphi$ is
$k$-positive for all extensions, i.e. for $k=2,3,\ldots$, then
$\varphi$ is completely positive (CP). Actually, it was shown by
Choi that $\varphi : M_d \longrightarrow M_d$ is CP iff it is
$d$-positive. Note, that using the hierarchy of cones
$\mathrm{V}_k$ we may reformulate the above definitions as
follows: a linear map $\varphi$ is $k$-positive iff
\begin{equation}\label{}
    (\oper_d \ot \varphi)(\mathrm{V}_k) \subset
    (M_d \ot M_d)^+\ .
\end{equation}

Let us denote by $\mathrm{P}_k$ a convex cone of $k$-positive
maps. One has, therefore, a natural chain of inclusions
\begin{equation}\label{P-k}
\mathrm{P}_d \subset \mathrm{P}_{d-1} \subset \ldots \subset
\mathrm{P}_2 \subset \mathrm{P}_1 \ ,
\end{equation}
where $\mathrm{P}_d$ stands for CP maps.
 Due to the celebrated Kraus theorem  any CP map
can be written in the following  Kraus representation
\begin{equation}\label{Kraus}
    \varphi(a) = \sum_\alpha\, K_\alpha a K^\dag_\alpha\ ,
\end{equation}
with $K_\alpha \in M_d$. Additional condition $\sum_\alpha
K^\dag_\alpha K_\alpha = I_d$ implies that $\mbox{Tr}\,\varphi(a)
= \mbox{Tr}\, a$.

Note, that we cannot detect entangled state using CP map.
Therefore, we are interested in positive maps which are not CP. It
turns out that  any positive map $\varphi$ may be written as a
difference of two CP maps, i.e.
\begin{equation}\label{Kraus}
    \varphi(a) = \sum_\alpha\, K_\alpha a K^\dag_\alpha -
    \sum_\beta\, L_\beta a L^\dag_\beta\ ,
\end{equation}
with $K_\alpha, L_\beta \in M_d$. The most prominent example of a
positive map which is not completely positive is a transposition
$\tau (a)= a^T$. Composing positive maps with transposition gives
rise to a new class of maps: a map $\varphi : M_d \longrightarrow
M_d$ is called $k$-copositive iff $\varphi \circ \tau$ is
$k$-positive. Finally, $\varphi$ is completely copositive (CcP)
iff $\varphi \circ \tau$ is CP. Equivalently, $\varphi$ is
$k$-copositive iff
\begin{equation}\label{}
    (\oper_d \ot \varphi)(\mathrm{V}^k) \subset
    (M_d \ot M_d)^+\ .
\end{equation}
Denoting by $\mathrm{P}^k$ a convex cone of $k$-copositive maps
one has
\begin{equation}\label{P=k}
\mathrm{P}^d \subset \mathrm{P}^{d-1} \subset \ldots \subset
\mathrm{P}^2 \subset \mathrm{P}^1 \ ,
\end{equation}
where $\mathrm{P}^d$ stands for CcP maps.

A crucial role in detecting quantum entanglement is played by
indecomposable maps: a positive map $\varphi$ is decomposable iff
it can be written as $\varphi = \varphi_1 + \varphi_2$ with
$\varphi_1$  and $\varphi_2$ being CP and CcP maps, respectively.
 Otherwise it is called indecomposable. Note that a
positive partial transpose (PPT) state can not be detected by any
decomposable map. Therefore, to detect PPT entangled states one
needs indecomposable maps. Having defined cones $\mathrm{P}_r$ and
$\mathrm{P}^s$ let $\mathrm{P}_r + \mathrm{P}^s$ stand for a set
of maps which can be written as $\varphi = \varphi_1 + \varphi_2$
with $\varphi_1 \in \mathrm{P}_r$  and $\varphi_2\in
\mathrm{P}^s$. Clearly, $\varphi$ is indecomposable iff $\varphi
\notin \mathrm{P}_d + \mathrm{P}^d$. An important subset  of
indecomposable maps contains so called atomic ones
\cite{Tomiyama3}: $\varphi$ is atomic iff $\varphi \notin
\mathrm{P}_2 + \mathrm{P}^2$. The importance of atomic maps
follows from the fact that they may be used to detect the
`weakest' bound entanglement.

Now,   $M_d \ot M_d$ is isomorphic to the space of linear maps
$\varphi : M_d \rightarrow M_d$ denoted by $\mathcal{L}(M_d,M_d)$:
for any $\varphi \in \mathcal{L}(M_d,M_d)$ one defines \cite{Jam}
\begin{equation}\label{}
\widehat{\varphi} = (\oper_d \ot \varphi) P^+ \in M_d \ot M_d\ ,
\end{equation}
where $P^+$ stands for (unnormalized) maximally entangled state in
$\Cd \ot \Cd$. If $e_i=|i\>$ $(i=1,\ldots,d)$ is an orthonormal
base in $\Cd$, then
\begin{equation}\label{J}
\widehat{\varphi}      = \sum_{i,j=1}^d e_{ij} \ot
\varphi(e_{ij})\ ,
\end{equation}
where $e_{ij} = |i\>\<j|$. Conversely, if $W\in M_d \ot M_d$ the
corresponding linear map is defined as follows
\begin{equation}\label{}
    \varphi_W(a) = \mbox{Tr}_2\left[ W \left(I_d \ot a^{\!{\rm T}}\right)\right] \
    .
\end{equation}
It is clear that if $\varphi$ is a positive but not CP map then
the corresponding operator $\widehat{\varphi}$ is an entanglement
witness.

Now, the space $\mathcal{L}(M_d,M_d)$ is endowed with a natural
inner product:
\begin{equation}\label{}
    (\varphi,\psi) = \mathrm{Tr} \Big( \sum_{\alpha=1}^{d^2}\,
    \varphi(f_\alpha)^\dag \psi(f_\alpha) \Big)\ ,
\end{equation}
where $f_\alpha$ is an arbitrary orthonormal base in $M_d$. Taking
$f_\alpha = e_{ij}$ one finds
\begin{eqnarray}\label{}
    (\varphi,\psi) &=& \mathrm{Tr} \Big( \sum_{i,j=1}^{d}\,
    \varphi(e_{ij})^\dag \psi(e_{ij}) \Big)\nonumber \\ & =& \mathrm{Tr} \Big( \sum_{i,j=1}^{d}\,
    \varphi(e_{ij})\psi(e_{ji}) \Big)\ .
\end{eqnarray}
This inner product is compatible with the standard Hilbert-Schmidt
product in $M_d \ot M_d$. Indeed,  taking $\widehat{\varphi}$ and
$\widehat{\psi}$ corresponding to $\varphi$ and $\psi$, one has
\begin{equation}\label{}
    (\widehat{\varphi},\widehat{\psi})_{\rm HS} =  \mathrm{Tr} (\widehat{\varphi}^\dag\widehat{\psi})
\end{equation}
and using (\ref{J}) one easily finds
\begin{equation}\label{}
(\varphi,\psi) = (\widehat{\varphi},\widehat{\psi})_{\rm HS}\ ,
\end{equation}
that is, formula (\ref{J}) defines an inner product isomorphism.
This way one establishes the duality between maps from
$\mathcal{L}(M_d,M_d)$ and operators from $M_d \ot M_d$
\cite{Eom}: for any $\rho \in M_d \ot M_d$ and $\varphi \in
\mathcal{L}(M_d,M_d)$ one defines
\begin{equation}\label{}
\< \rho, \varphi\> = (\rho,\widehat{\varphi})_{\rm HS} \ .
\end{equation}
In particular, if $\rho$ is an unnormalized state and $\varphi$ is
a positive map, then
\begin{equation}\label{dual}
\< \rho, \varphi\> = \mathrm{Tr}(\widehat{\varphi} \rho) =
\mathrm{Tr} \Big( \sum_{i,j=1}^{d}\,
    \varphi(e_{ij})\,\rho_{ji} \Big)\ ,
\end{equation}
where
\begin{equation}\label{}
    \rho = \sum_{i,j=1}^d\, e_{ij} \ot \rho_{ij} \ ,
\end{equation}
with $\rho_{ij} \in M_d$. Formula (\ref{dual}) reproduces the
formula for an entanglement witness $W =\widehat{\varphi}$.

This construction shows that two sets of cones --- $\mathrm{V}_k$
and $\mathrm{P}^k$ --- are dual to each other. It follows from
(\ref{dual}) that
\[ \rho \in \mathrm{V}_r \ \ \Longleftrightarrow\ \ \<
\rho,\varphi\> \geq 0\  \ \ \mathrm{for\ all}\ \ \ \varphi \in
\mathrm{P}^r\ .
\]
Moreover,
\[ \rho \in \mathrm{V}_r\cap \mathrm{V}^s \ \ \Longleftrightarrow\ \ \<
\rho,\varphi\> \geq 0\  \ \ \mathrm{for\ all}\ \ \ \varphi \in
\mathrm{P}^r + \mathrm{P}_s\ .
\]
Conversely,
\[ \varphi \in \mathrm{P}^r \ \ \Longleftrightarrow\ \ \<
\rho,\varphi\> \geq 0\  \ \ \mathrm{for\ all}\ \ \ \rho \in
\mathrm{V}_r\ ,
\]
and
\[ \varphi \in \mathrm{P}^r + \mathrm{P}_s  \ \ \Longleftrightarrow\ \ \<
\rho,\varphi\> \geq 0\  \ \ \mathrm{for\ all}\ \ \ \varphi \in
\mathrm{V}_r\cap \mathrm{V}^s\ .
\]

Clearly, formula (\ref{dual}) may be used to witness entanglement:
 $\rho$ is entangled iff there exists $\varphi \in P^1$ such that
 $\< \rho,\varphi\> <0$. More generally, a positive operator $\rho
 \notin \mathrm{V}_r$ iff there exists $\varphi \in P^r$ such that
 $\< \rho,\varphi\> <0$, and $\rho
 \notin \mathrm{V}_r \cap \mathrm{V}^s$ iff there exists $\varphi \in P^r + \mathrm{P}_s$ such that
 $\< \rho,\varphi\> <0$.

Dually, we may use (\ref{dual}) to check whether a given positive
map $\varphi$ is indecomposable or atomic: $\varphi$ is
indecomposable iff there exists $\rho \in \mathrm{V}_d \cap
\mathrm{V}^d$ (i.e. $\rho$ is PPT) such that
 $\< \rho,\varphi\> <0$. Finally, $\varphi$ is
atomic iff there exists $\rho \in \mathrm{V}_2 \cap \mathrm{V}^2$
such that $\< \rho,\varphi\> <0$.

\section{Indecomposable maps -- review}
\label{REVIEW}

For $d=2$ all positive maps $\varphi : M_2 \rightarrow M_2$  are
decomposable \cite{Woronowicz1,Woronowicz2}.

\subsection{Choi map for $d=3$}

The first example of an indecomposable positive linear map in
$M_3$ was found by Choi \cite{Choi1}. The (normalized) Choi map
reads as follows
\begin{eqnarray}\label{Choi-map}
    \Phi_{\rm C}(e_{ii}) &=& \sum_{i,j=1}^3 a^{\rm
    C}_{ij} e_{jj} \ , \nonumber\\
\Phi_{\rm C}(e_{ij}) &=& - \frac 12 \, e_{ij} \ , \ \ \ \ i\neq j\
,
\end{eqnarray}
where $[a^{\rm C}_{ij}]$ is the following bistochatic matrix:
\begin{equation}\label{a-Choi}
    a^{\rm C}_{ij} = \frac 12 \left( \begin{array}{ccc} 1 & 1 &
    0 \\ 0 & 1 & 1 \\
    1 & 0 & 1  \end{array} \right) \ .
\end{equation}
This map may be generalized as follows \cite{Cho-Kye}: for any
$a,b,c\geq 0$ let us define
\begin{eqnarray}\label{Choi-map}
    \Phi[a,b,c](e_{ii}) &=& \sum_{i,j=1}^3 a_{ij} e_{jj} \ , \nonumber\\
\Phi[a,b,c](e_{ij}) &=& - \frac{1}{a +b+c} \ e_{ij} \ , \ \ i\neq
j\ ,
\end{eqnarray}
with
\begin{equation}\label{}
    a_{ij} = \frac{1}{a +b+c}\, \left( \begin{array}{ccc} a & b &
    c \\ c & a & b \\
    b & c & a  \end{array} \right) \ .
\end{equation}
Clearly, $\Phi^{\rm C}=\Phi[1,1,0]$. The map $\Phi[1,0,\mu]$ with
$\mu \geq 1$ is the example of indecomposable map introduced by
Choi \cite{Choi2}. Now, it was shown \cite{Cho-Kye} that
$\Phi[a,b,c]$ is an indecomposable positive map if and only if the
following conditions are satisfied:
\begin{eqnarray*}\label{}
&(i)& \ \ 0 \leq a < 2 \ ,\\
&(ii)& \ \ a+b+c\geq 2 \ , \\
&(iii)& \left\{ \begin{array}{ll} (1-a)^2 \leq bc < (2-a)^2/4 \ , &  \ \mbox{if}\ \ 0 \leq a \leq 1\\
0 \leq bc < (2-a)^2/4 \ , & \ \mbox{if}\ \ 1 \leq a < 2
\end{array} \right.\ .
\end{eqnarray*}
Actually, $\Phi[a,b,c]$ is indecomposable if and only if it is
atomic, i.e. it cannot be decomposed into the sum of a 2-positive
and 2-copositive maps.

\subsection{Indecomposable  maps for $d\geq 3$}

 For $d\geq 3$ there are three basic
 families of indecomposable maps:

1) A discrete family $\tau_{d,k}$, $k=1,\ldots, d-2$ \cite{Ha1}.
Let $s$ be a unitary shift defined by:
\[  s\, e_i = e_{i+1} \ ,\ \ \ \ \ i=1, \ldots , d\ , \]
where the indices are understood mod $d$. The maps $\tau_{d,k}$
are defined as follows:
\begin{equation}\label{}
    \tau_{d,k}(X) = (d-k)\, \epsilon(X) + \sum_{i=1}^k\,
    \epsilon(s^i\, X\, {s^{* i}}) - X \ ,
\end{equation}
where  $\epsilon(X)$ is defined in (\ref{ep}). The map
$\tau_{d,0}$ defined in (\ref{d0}) is completely positive and it
is well known that the map corresponding to $k=d-1$ is completely
co-positive \cite{Ha1}.

Note that $\tau_{d,k}(I_d) = (d-1)I_d$, and
$\mbox{Tr}\,\tau_{d,k}(X) = (d-1)\mbox{Tr}\,X$, hence the
normalized maps
\begin{equation}\label{Phi-d-k}
    \Phi_{d,k}(X) = \frac{1}{d-1}\, \tau_{d,k}(X) \ ,
\end{equation}
are bistochastic. In particular $\Phi[1,0,1] = \Phi_{3,1}$.

2) A class of maps $\varphi_{\bf p}$ parameterized by $d+1$
parameters ${\bf p} = (p_0,p_1,\ldots,p_d)$:
\begin{eqnarray}\label{}
\varphi_{\bf p}(e_{11}) &=& p_0e_{11} + p_de_{dd}\ , \nonumber\\
\varphi_{\bf p}(e_{22}) &=& p_0e_{22} + p_1e_{11}\ , \nonumber\\
 &\vdots&  \\
\varphi_{\bf p}(e_{dd}) &=& p_0e_{dd} + p_{d-1}e_{d-1,d-1}\ , \nonumber\\
\varphi_{\bf p}(e_{ij}) &=& -e_{ij}\ ,\ \ \ i\neq j\ . \nonumber
\end{eqnarray}
It was shown \cite{Osaka2,Ha2} that if
\begin{eqnarray*}\label{}
&a)& \ \  p_1,\ldots,p_d>0\ , \\
&b)& \ \  d-1 > p_0 \geq d-2 \ , \\
&c)&\ \  p_1 \cdot \ldots \cdot p_d \geq (d-1-p_0)^d\ ,
\end{eqnarray*}
then $\varphi_{\bf p}$ is a positive indecomposable map. Actually,
$\varphi_{\bf p}$ is atomic, i.e. it cannot be decomposed into the
sum of a 2-positive and 2-copositive maps.

3) A family of maps constructed by Terhal \cite{Terhal2} from
unextendible product bases \cite{UPB,UPB-inne}. Let $|\alpha_i\>
\ot |\beta_i\>$; $i=1,\ldots,K < d^2$ be an unextendible product
basis in $\Cd \ot \Cd$. Then an unnormalised density matrix
\begin{equation*}\label{}
    \rho =   I_d \ot I_d - \sum_{i=1}^K |\alpha_i\>\<\alpha_i| \ot
    |\beta_i\>\<\beta_i| \ ,
\end{equation*}
defines a PPT entangled state. This state my be detected by the
following entanglement witness:
\begin{equation}\label{}
    W = \sum_{i=1}^K |\alpha_i\>\<\alpha_i| \ot
    |\beta_i\>\<\beta_i| - d\varepsilon|\Psi\>\<\Psi| \ ,
\end{equation}
where $|\Psi\>$ is a maximally entangled state such that
$\<\Psi,\rho \Psi\> > 0$. A parameter $\varepsilon$ is defined by
\begin{equation}\label{}
    \varepsilon = \min_{|\phi_1\> \ot |\phi_2\>} \, \sum_{i=1}^K |\<\alpha_i|\phi_1\>|^2
    \<\beta_i|\phi_2\>|^2\ ,
\end{equation}
where the minimum is taken over all pure separable states
$|\phi_1\> \ot |\phi_2\>$. It is therefore clear that the
corresponding map
\begin{equation}\label{}
    \Phi(X) = \mbox{Tr}_2\left[ W \left(I_d \ot X^{\!{\rm T}}\right)\right] \ ,
\end{equation}
is an indecomposable positive map in $M_d$.

Another example of an indecomposable map (also outside the class
(\ref{GEN})) was given by Robertson
\cite{Robertson1,Robertson2,Robertson3,Robertson4}. Robertson map
$\varphi_4 : M_4 \rightarrow M_4$ is defined by
\begin{eqnarray*}\label{}
    \varphi_4(e_{11}) &=&  \varphi_4(e_{22}) = \frac 12 ( e_{33} +
    e_{44}) \ , \\
 \varphi_4(e_{33}) &=&  \varphi_4(e_{44}) = \frac 12 ( e_{11} +
    e_{22}) \ , \\
  \varphi_4(e_{13}) &=& \frac 12 (e_{13} + e_{42})\ , \\
\varphi_4(e_{14}) &=& \frac 12 (e_{14} - e_{32})\ , \\
\varphi_4(e_{23}) &=& \frac 12 (e_{23} - e_{41})\ , \\
\varphi_4(e_{24}) &=& \frac 12 (e_{24} + e_{31})\ ,
\end{eqnarray*}
and the remaining
\[  \varphi_4(e_{12}) =  \varphi_4(e_{21}) =  \varphi_4(e_{34})
 =  \varphi_4(e_{43}) = 0\ . \]
It satisfies $\varphi_4(I_4)=I_4$ and $\mbox{Tr}\, \varphi_4(X) =
\mbox{Tr}\, X$, and it is known that $\varphi_4$ is atomic and
hence indecomposable.

\section{On certain class of positive maps}
\label{CLASS}

Consider the following class of linear maps $\varphi : M_d
\longrightarrow M_d$:
\begin{eqnarray}
    \varphi(e_{ii})  &=& \sum_{j=1}^d a_{ij} e_{jj} \ ,\nonumber \\ \label{GEN}
    \varphi(e_{ii})  &=& - e_{ij} \ , \ \ \ \ i\neq j \ ,
\end{eqnarray}
with $||a_{ij}||$ being a $d \times d $  real positive matrix. Let
us observe that most of well known positive maps reviewed in the
previous section do belong to this class (only the class based on
unextendible product bases and the example constructed by
Robertson do not).

\begin{theorem} \label{TH1}
A map belonging to a class (\ref{GEN}) is positive iff
\begin{equation}\label{B>0}
     \left( 1 - \sum_{i=1}^d\,
    \frac{|x_i|^2}{B_{i}(x)} \right) \prod_{k=1}^d B_{k}(x) \geq 0\ ,
\end{equation}
for all $x \in \mathbb{C}^d$ such that $|x|^2 = \sum_{i=1}^d
|x_i|^2=1$, and
\begin{equation}\label{Bii}
    B_{i}(x) =  |x_i|^2 + \sum_{j=1}^d\, a_{ij} |x_j|^2 \ .
\end{equation}
If all $B_{i}\neq 0$, then (\ref{B>0}) simplifies to
\begin{equation}\label{B>0s}
    \sum_{i=1}^d\, \frac{|x_i|^2}{B_{i}(x)} \leq 1\ .
\end{equation}
\end{theorem}
{\it Proof:} $\varphi$ is positive iff for any normalized $x \in
\mathbb{C}^d$ one has $\varphi(P_x) \geq 0$, where $P_x =
|x\>\<x|$ denotes the corresponding 1-dimensional projector. Let
us denote the corresponding $d \times d$ matrix $\varphi(P_x)$ by
$A(x) = [A_{ij}(x)]$, that is
\begin{eqnarray}\label{}
    A_{ii}(x) &=& \sum_{j=1}^d a_{ij} |x_j|^2 \ , \\
    A_{ij}(x) &=& - x_i \overline{x}_j \ , \ \ \ \ i\neq j \ .
\end{eqnarray}
Positivity of $\varphi$ is therefore equivalent to the positivity
of $A(x)$ for any normalized $x \in \mathbb{C}^d$. Now, to check
for positivity of $A(x)$ one computes the characteristic
polynomial
\begin{equation}\label{det0}
    \mbox{det}\, ||\, A_{ij}(x) - \lambda\delta_{ij}\, || = \sum_{k=0}^d
    (-\lambda)^{d-k} C_k(x) \ ,
\end{equation}
and $||A_{ij}(x)|| \geq 0$ iff $C_k(x) \geq 0$ for
$k=0,1,\ldots,d$. The determinant  of $|| A_{ij}(x) -
\lambda\delta_{ij} ||$ is easy to calculate. Using the following
formula
\begin{equation}\label{DET}
     \left| \begin{array}{ccccc} \gamma_1 & \alpha_2
    \beta_1 & \alpha_3\beta_1 & \ldots & \alpha_n \beta_1 \\
 \alpha_1\beta_2 &\gamma_2 & \alpha_3\beta_2 & \ldots & \alpha_n \beta_2 \\
 \vdots &  \vdots & \vdots & \ddots & \vdots \\
 \alpha_1\beta_n & \alpha_2\beta_n & \alpha_3\beta_n & \ldots &
 \gamma_n \end{array} \right| \ = \ \left( 1 + \sum_{k=1}^n
 \frac{\alpha_k\beta_k}{\gamma_k - \alpha_k\beta_k} \right) \prod_{i=1}^n(\gamma_i -
 \alpha_i\beta_i)\ ,
\end{equation}
with  $\gamma_k \neq \alpha_k\beta_k$ for $ k=1,2,\ldots,n$, one
easily finds
\begin{equation}\label{det1}
\mbox{det}\, ||\, A_{ij}(x) - \lambda\delta_{ij}\, || \ = \ \left(
1 - \sum_{k=1}^d
 \frac{|x_k|^2}{ B_{k}(x)  -\lambda} \right) \prod_{i=1}^d( B_{i}(x)
 - \lambda)\ ,
\end{equation}
where $B_{k}(x)$ is given by (\ref{Bii}). Now, formula
(\ref{det0}) implies for the coefficients $C_{d-l}(x)$
\begin{equation}\label{}
    C_{d-l}(x) = \frac{(-1)^l}{(l-1)!}\,
    \frac{d^l}{d\lambda^l}\,  \mbox{det}\, ||\, A_{ij}(x) - \lambda\delta_{ij}\,
    ||\, \Big|_{\lambda = 0}\ ,
\end{equation}
and hence, using (\ref{det1}) one finds
\begin{equation}\label{Ck}
    C_k(x) = \left( 1 - \sum_{i=1}^d\,
    \frac{|x_i|^2}{B_{i}(x)} \right) \sum_{i_1 < i_2 < \ldots < i_k} B_{i_1}(x) \ldots
    B_{i_k}(x) \ .
\end{equation}
It is therefore clear that $C_k(x)\geq 0$ iff $C_d(x)\geq$. Hence,
using (\ref{Ck}) one obtains the following condition for the
positivity of $\varphi$:
\begin{equation}\label{}
    C_d(x) =  \left( 1 - \sum_{i=1}^d\,
    \frac{|x_i|^2}{B_{i}(x)} \right) \prod_{k=1}^d B_{k}(x) \geq 0\ ,
\end{equation}
which finally proves (\ref{B>0}). \hfill $\Box$

As a direct application of  Theorem \ref{TH1} let us observe that
a celebrated Choi map in $M_3$ defined by the matrix
\begin{equation}\label{}
a_{ij} \ = \  \left( \begin{array}{ccc} 1 & 1 &
    0 \\ 0 & 1 & 1 \\
    1 & 0 & 1  \end{array} \right) \ ,
\end{equation}
gives rise to
\begin{eqnarray*}\label{}
    B_{1}(x) &=& 2|x_1|^2  + |x_2|^2\ , \\
B_{2}(x) &=& 2|x_2|^2  + |x_3|^2\ , \\
B_{3}(x) &=& 2|x_3|^2  + |x_1|^2\ .
\end{eqnarray*}
and direct calculation shows that the condition (\ref{B>0}) is
satisfied and hence $\varphi$ is positive.

Let us observe that the map $\varphi$ defined in (\ref{GEN})
acting on $X \in M_d$ give
\begin{eqnarray}\label{}
    \varphi(X) &=& \sum_{i,j=1}^d (a_{ij} + \delta_{ij})\, e_{ii}
    \<e_j|Xe_j\> - X \nonumber \\ &=&
\sum_{i,j=1}^d (a_{ij} + \delta_{ij})\, e_{ij} \, X\, e_{ij}^*
     - \sum_{i,j=1}^d e_{ii}\,X\, e_{jj} \nonumber \ .
\end{eqnarray}
Now introducing $||b_{ij}||$ by
\begin{equation}\label{}
    b_{ii} = a_{ii} \ , \ \ \ \ b_{ij} = -1\ , \ \ i\neq j\ ,
\end{equation}
one has
\begin{equation}\label{}
 \varphi(X) = \sum_{i\neq j}^d a_{ij} \, e_{ij} \, X\, e_{ij}^*
     + \sum_{i\neq j}^d b_{ij} \,  e_{ii}\,X\, e_{jj} \ .
\end{equation}
This observation gives rise to the following
\begin{theorem}
A map  $\varphi$ defined in (\ref{GEN}) is completely positive iff
the matrix $||b_{ij}||$ is positive.
\end{theorem}
Let us note that if $a_{11} = \ldots = a_{dd} = a$, that is,
\begin{equation}\label{}
 b_{ij} = \left( \begin{array}{ccccc} a & -1 & -1 & \ldots & -1 \\
 -1 & a & -1 & \ldots & -1 \\
 \vdots & \vdots& \vdots & \ddots & \vdots \\
 -1 & -1 & -1 & \ldots & a \end{array} \right) \ ,
\end{equation}
then $||b_{ij}||\geq 0$ iff $a\geq d-1$. In particular if $a_{ij}
= a \delta_{ij}$ then the action of $\varphi$ is given by
\begin{equation}\label{phi-a+1}
    \varphi(X) = (a+1)\epsilon(X) - X \ ,
\end{equation}
where  $\epsilon : M_d \longrightarrow M_d$ stands for the
projector onto the diagonal part:
\begin{equation}\label{ep}
 \epsilon(X) = \sum_{i=1}^d  \mbox{Tr}[X\,e_{ii}]\, e_{ii}\ .
\end{equation}
Hence the map (\ref{phi-a+1}) is completely positive for $a\geq
d-1$. For $a=d-1$ one recovers a CP map
\begin{equation}\label{d0}
    \tau_{d,0}(X) = d\,\epsilon(X) - X \ ,
\end{equation}
considered in \cite{Ha1}.

\section{A family of positive maps parameterized by contractions}
\label{CONTRACTIONS}

Recently a rich family of positive maps was constructed in
\cite{Kossak1}. Let us consider the following map $\varphi : M_d
\longrightarrow M_d$:
\begin{equation}\label{Kossak-ball}
    \varphi(X) = \frac{I_d}{d}\, \mbox{Tr}X + \frac{1}{d-1}\,
    \sum_{\alpha,\beta=1}^{d^2-1} f_\alpha A_{\alpha\beta}
    \mbox{Tr}(f_\beta X) \ ,
\end{equation}
where $A=[A_{\alpha\beta}]$ is a real matrix representing
contraction in $\mathbb{R}^{d^2-1}$ and $f_\alpha \in M_d$ define
the generators of $SU(d)$ such that $f_\alpha = f_\alpha^*$,
$\mbox{Tr}(f_\alpha f_\beta)=\delta_{\alpha\beta}$, and
$\mbox{Tr}\, f_\alpha =0$ for $\alpha,\beta = 1, \ldots , d^2-1$.
 The explicit construction of $f_\alpha$ reads
as follows:
\begin{equation*}\label{}
    (f_1, \ldots , f_{d^2-1}) = (d_l,u_{kl},v_{kl})\ , \ \
\end{equation*}
for $l=1,\ldots , d-1$ and  $1\leq k < l \leq d$, where the
diagonal operators
\begin{eqnarray}\label{d}
d_l &=&  \frac{1}{\sqrt{l(l+1)}}\, \Big( \sum_{k=1}^l e_{kk} -
le_{l+1,l+1} \Big) \ ,
\end{eqnarray}
define Cartan subalgebra of $su(d)$, and off-diagonal
\begin{eqnarray}\label{off-d}
    u_{kl} = \frac{1}{\sqrt{2}} (e_{kl} + e_{lk}) \ , \ \ \ \
    v_{kl} = \frac{-i}{\sqrt{2}} (e_{kl} - e_{lk}) \ .
\end{eqnarray}
It was shown in \cite{Kossak1} that this map is positive for an
arbitrary contraction $A_{\alpha\beta}$. Moreover, one has
\[ \varphi(I_d)=I_d \ \ \  {\rm and}\ \ \  \mbox{Tr}\, \varphi(X) =
\mbox{Tr}\, X\ . \]

Consider now a special case corresponding to
\begin{equation}\label{}
   A = \left( \begin{array}{cc} \mathbf{A} & 0 \\ 0 & -
    I    \end{array} \right) \ ,
\end{equation}
where $\mathbf{A}$ is a contraction in $\mathbb{R}^{d-1}$. Recall,
that any contraction $\mathbf{A}_{\alpha\beta}$ may be represented
as follows
\begin{equation}\label{ARR}
    \mathbf{A} = R_1 D R_2\ ,
\end{equation}
where $R_i$ represent rotations in $\mathbb{R}^{d-1}$, i.e. $R_i
\in SO(d-1)$, and $D$ is a diagonal matrix with $|\lambda_i| =
|D_{ii}|\leq 1$. Let us consider the special case of (\ref{ARR})
such that $D=\lambda I_{d-1}$ with $0\leq \lambda \leq 1$, that
is,
\begin{equation}\label{}
    \mathbf{A} = \lambda\, R\ ,
\end{equation}
where $R = R_1R_2 \in SO(d-1)$. The general case (\ref{ARR})
produces much more complicated situation  even in $d=3$ (see
Appendix).
 The action of $\varphi$ is given by
\begin{eqnarray}\label{}
\varphi(e_{ij}) &=& - \frac{1}{d-1}\, e_{ij} \ , \ \ \ i\neq j \ ,
\label{phi-eij}\\
\varphi(e_{ii}) &=& \frac{I_d}{d} + \frac{\lambda}{d-1}\,
    \sum_{\alpha,\beta=1}^{d-1} f_\alpha {R}_{\alpha\beta}
(e_i,f_\beta e_i)\ . \label{phi-eii}
\end{eqnarray}
Note that $\varphi(e_{ii})$ may be rewritten as follows
\begin{equation}\label{a-bi}
\varphi(e_{ii}) = \sum_{j=1}^d a_{ij} e_{jj} \ ,
\end{equation}
where $a_{ij} = \mbox{Tr}[\varphi(e_{ii})e_{jj}]$ is given by the
following bistochastic matrix:
\begin{equation}\label{a-R}
    a_{ij} = \frac 1d + \frac{\lambda}{d-1} \sum_{\alpha,\beta=1}^{d-1}
    (e_j,f_\alpha e_j) {R}_{\alpha\beta}
(e_i,f_\beta e_i)\ .
\end{equation}
Therefore, up to the normalization factor $1/(d-1)$, this family
belongs to our class (\ref{GEN}) discussed in the previous
section. Consider now a class of positive maps defined by
\begin{eqnarray}
    \varphi(e_{ii})  &=& \sum_{j=1}^d a_{ij} e_{jj} \ ,\nonumber \\ \label{GEN-b}
    \varphi(e_{ii})  &=& - \frac{1}{d-1}\,e_{ij} \ , \ \ \ \ i\neq j \ ,
\end{eqnarray}
with bistochastic $||a_{ij}||$.

\begin{theorem}
A bistochastic matrix $||a_{ij}||$ corresponds to contraction
$\lambda R$ with $R\in SO(d-1)$ and $\lambda\leq 1$, that is
$||a_{ij}||$ is given by (\ref{a-R}),  iff
\begin{equation}\label{aa}
    \sum_{k=1}^d\, a_{ik}a_{jk}\ =\ \frac{1}{(d-1)^2}\, \left( \lambda^2\delta_{ij}
    + d-2 + \frac{1-\lambda^2}{d} \right) \ .
\end{equation}
\end{theorem}
{\it Proof:} To prove (\ref{aa}) define a new map $\Phi : M_d
\longrightarrow M_d$
\begin{eqnarray}\label{A1}
    \Phi(e_{ii}) = (d-1) \left( \varphi(e_{ii}) - \frac{I_d}{d}
    \right) = \lambda \sum_{\alpha,\beta=1}^{d-1} f_\alpha {R}_{\alpha\beta}
(e_i,f_\beta e_i)\ ,
\end{eqnarray}
together with a dual map
\begin{eqnarray}\label{A2}
    \widetilde{\Phi}(e_{ii}) = (d-1) \left( \widetilde{\varphi}(e_{ii}) - \frac{I_d}{d}
    \right) = \lambda \sum_{\alpha,\beta=1}^{d-1} (e_i,f_\alpha e_i)
    {R}_{\alpha\beta}f_\beta
\ .
\end{eqnarray}
One has $  \Phi(I_d)=\widetilde{\Phi}(I_d)=0$. Now, let us compute
$\widetilde{\Phi}[\Phi(e_{ii})] $:
\begin{eqnarray}\label{}
\widetilde{\Phi}[\Phi(e_{ii})] = \lambda
\sum_{\alpha,\beta=1}^{d-1} \widetilde{\Phi}(f_\alpha)
{R}_{\alpha\beta} (e_i,f_\beta e_i) = \lambda
\sum_{\alpha,\beta=1}^{d-1}\sum_{j=1}^d
\widetilde{\Phi}(e_{jj})(e_j,f_\alpha e_j) {R}_{\alpha\beta}
(e_i,f_\beta e_i)\ ,
\end{eqnarray}
and hence using (\ref{A2})
\begin{eqnarray}\label{}
\widetilde{\Phi}[\Phi(e_{ii})] \ = \
\lambda^2\sum_{\alpha,\beta=1}^{d-1}\,
\sum_{\mu,\nu=1}^{d-1}\sum_{j=1}^d (e_j,f_\alpha e_j)
{R}_{\alpha\beta} (e_i,f_\beta e_i) (e_j,f_\mu
e_j){R}_{\mu\nu}f_\nu \ .
\end{eqnarray}
Taking into account that
\[   \sum_{j=1}^d (e_j,f_\alpha e_j)(e_j,f_\mu e_j) =
\delta_{\alpha\mu}\ ,  \] one obtains
\begin{eqnarray}\label{}
\widetilde{\Phi}[\Phi(e_{ii})] &=& \lambda^2
\sum_{\alpha,\beta,\nu=1}^{d-1}
{R}_{\alpha\beta}{R}_{\alpha\nu}(e_i,f_\beta e_i)f_\nu = \lambda^2
\sum_{\beta=1}^{d-1}(e_i,f_\beta e_i)f_\beta\nonumber \\ & =&
\lambda^2 \left( \sum_{\beta=0}^{d-1}(e_i,f_\beta e_i)f_\beta -
(e_i,f_0 e_i)f_0 \right) \ ,
\end{eqnarray}
where $f_0 = I_d/\sqrt{d}$. This leads to the following formula
\begin{equation}\label{A3}
\widetilde{\Phi}[\Phi(e_{ii})] = \lambda^2 \left( e_{ii} -
\frac{I_d}{d} \right) \ .
\end{equation}
Now, using (\ref{A1})
\begin{eqnarray*}\label{}
\widetilde{\Phi}[\Phi(e_{ii})] &=&  (d-1)\widetilde{\Phi}\left[
\varphi(e_{ii}) - \frac{I_d}{d}\right] \ = \  (d-1)\sum_{j=1}^d
a_{ij} \widetilde{\Phi}(e_{jj}) \nonumber \\ &=&
(d-1)^2\sum_{j=1}^d a_{ij} \left[ \widetilde{\varphi}(e_{jj}) -
\frac{I_d}{d} \right] \  = \  (d-1)^2 \left[ \sum_{j=1}^d a_{ij}\,
\widetilde{\varphi}(e_{jj}) - \frac{I_d}{d} \right]\ ,
\end{eqnarray*}
where we have used $\sum_{j=1}^d a_{ij} = 1$. Finally, taking into
account the definition of the dual map
\[  \widetilde{\varphi}(e_{jj}) = \sum_{k=1}^d a_{kj} e_{kk}\ , \]
one gets
\begin{eqnarray}\label{A4}
\widetilde{\Phi}[\Phi(e_{ii})] = (d-1)^2 \left[ \sum_{j,k=1}^d
a_{ij}\,a_{kj}e_{kk} - \frac{I_d}{d} \right]\, ,
\end{eqnarray}
and comparing formulae (\ref{A3}) and (\ref{A4})
\[ (d-1)^2 \left( \sum_{j,k=1}^d
a_{ij}\,a_{kj}e_{kk} - \frac{I_d}{d} \right) = \lambda^2 \left(
e_{ii} - \frac{I_d}{d} \right)\ , \]
 one shows (\ref{aa}). \hfill $\Box$

\section{Main result}
\label{MAIN}

Now we show that for certain class of bistochastic satisfying
(\ref{aa}) matrices the corresponding positive map (\ref{GEN-b})
is indecomposable.

\begin{theorem}
Let $\varphi : M_d \longrightarrow M_d$ be a positive map defined
by (\ref{GEN-b}) with a bistochastic matrix $||a_{ij}||$
satisfying (\ref{aa}). Suppose that a matrix $\,||a_{ij}||$ is
cyclic, i.e.
\begin{equation}\label{}
    a_{ij} = \left( \begin{array}{ccccc} \alpha_0 & \alpha_1 &
    \alpha_2 & \ldots & \alpha_{d-1} \\ \alpha_{d-1} & \alpha_0 &
    \alpha_1 & \ldots & \alpha_{d-2} \\
    \vdots &\vdots & \vdots &\ddots &\vdots \\
  \alpha_1 & \alpha_2 &
    \alpha_3 & \ldots & \alpha_0 \end{array} \right) \ ,
\end{equation}
with $\alpha_i \geq 0$, and $\alpha_0 + \alpha_1 + \ldots +
\alpha_{d-1} =1$. Then $\varphi$ is indecomposable if :\\
 i) for $d=2k+1$ one  of the
following two conditions is satisfied
\begin{eqnarray*}\label{}
&1)&\ \   \left\{ \begin{array}{l}\alpha_1 + \ldots +\alpha_k
> 0
\\
\alpha_1 + \ldots +\alpha_k  \neq
    \alpha_{k+1} + \ldots +\alpha_{2k} \end{array} \right. \ , \\
&2)&\ \ \left\{ \begin{array}{l}\alpha_1 + \ldots +\alpha_k = 0
\\ 1 > \alpha_0 > 0
 \end{array} \right. \ ,
\end{eqnarray*}
ii) for $d=2k$  one  of the following two conditions is satisfied
\begin{eqnarray*}\label{}
&1)&\ \  \left\{ \begin{array}{l}\alpha_1 + \ldots +\alpha_{k-1}
> 0
\\
 \alpha_1 + \ldots +\alpha_{k-1}  \neq
    \alpha_{k+1} + \ldots +\alpha_{2k-1} \end{array} \right. \ , \\
&2)& \ \  \left\{ \begin{array}{l}\alpha_1 + \ldots +\alpha_{k-1}
= 0
\\ 1 > \alpha_0 +\alpha_k> 0
 \end{array} \right. \ .
\end{eqnarray*}

\end{theorem}
{\it Proof}. --- To show that a positive map $\varphi$ is
indecomposable we use the duality formula (\ref{dual}), i.e. we
construct a PPT matrix $\rho \in (M_d \ot M_d)^+$ such that $\<
\rho, \varphi\> < 0$. Consider the following matrix
\begin{equation}\label{}
    \rho = \sum_{i,j=1}^d A_{ij}\, e_{ij} \ot e_{ij} + \sum_{i\neq
    j} D_{ij}\, e_{ii} \ot e_{jj} \ .
\end{equation}
It is positive iff the Hermitian matrix $[A_{ij}]\geq 0$ and all
coefficients  $D_{ij}\geq 0$. It was shown in \cite{PPT-nasza}
that $\rho$ is PPT if
\begin{equation}\label{PPT-nasze}
    D_{ij}D_{ji} - |A_{ij}| \geq 0\ ,  \ \ \ \ i\neq j\ .
\end{equation}
Let us consider two separate cases:

i) If $d = 2k+1$, let us take $A_{ij} = a>0$, and
\begin{eqnarray*}\label{}
    D_{i,i+1} &=& D_{i,i+2} = \ldots = D_{i,i+k} = a^2\ , \\
    D_{i,i+k+1} &=& D_{i,i+k+2} = \ldots = D_{i,i+2k} = 1\ ,
\end{eqnarray*}
where  the addition is mod $d$. Clearly, the condition
(\ref{PPT-nasze}) is satisfied and hence the corresponding $\rho$
is PPT. Note, that $\<\rho,\varphi\>= dF(a)$ with
\begin{equation*}\label{}
    F(a) = -a(1-\alpha_0) + a^2( \alpha_1 + \ldots +\alpha_k)  +
    (\alpha_{k+1} + \ldots +\alpha_{2k}) \ .
\end{equation*}
Note, that if $\alpha_1 + \ldots + \alpha_k > 0$ the function
$F=F(a)$ attains its minimum for
\[  a=a_0 = \frac{1-\alpha_0}{2( \alpha_1 + \ldots +\alpha_k)} \ , \]
and
\begin{equation*}\label{}
    F(a_0) = -\frac{[ ( \alpha_1 + \ldots +\alpha_k) - (\alpha_{k+1} + \ldots
    +\alpha_{2k})]^2}{4( \alpha_1 + \ldots +\alpha_k)} \ ,
\end{equation*}
which, for  $ \alpha_1 + \ldots +\alpha_k \neq\alpha_{k+1} +
\ldots +\alpha_{2k}$, implies that $\< \rho,\varphi\> < 0$. Now,
if $\alpha_1 + \ldots + \alpha_k = 0$, then $\alpha_{k+1} + \ldots
+ \alpha_{2k} = 1-\alpha_0$ and
\[ F(a) = (1-a)(1-\alpha_0)  \ . \]
Hence, $F(a) < 0$ iff $a > 1$ and $1> \alpha_0 >0$.

ii) If $d = 2k$, let us take $A_{ij} = a>0$, and
\begin{eqnarray*}\label{}
    D_{i,i+1} &=& D_{i,i+2} = \ldots = D_{i,i+k-1} = a^2\ ,\\
D_{i,i+k} &=&a\ , \\
    D_{i,i+k+1} &=& D_{i,i+k+2} = \ldots = D_{i,i+2k-1} = 1\ .
\end{eqnarray*}
 Clearly, the condition (\ref{PPT-nasze}) is satisfied and hence
the corresponding $\rho$ is PPT. Note, that $\<\rho,\varphi\>=
dG(a)$ with
\begin{eqnarray*}\label{}
    G(a) &=& -a(1-\alpha_0-\alpha_k) + a^2( \alpha_1 + \ldots +\alpha_{k-1})
    \nonumber \\ &+&
    (\alpha_{k+1} + \ldots +\alpha_{2k-1}) \ .
\end{eqnarray*}
Now, if $\alpha_1 + \ldots +\alpha_{k-1}>0$ the function $G=G(a)$
attains its minimum for
\[  a=a_0' = \frac{1-\alpha_0-\alpha_k}{2( \alpha_1 + \ldots +\alpha_{k-1})} \ , \]
and
\begin{equation*}\label{}
    G(a_0') = -\frac{[ ( \alpha_1 + \ldots +\alpha_{k-1}) - (\alpha_{k+1} + \ldots
    +\alpha_{2k-1})]^2}{4( \alpha_1 + \ldots +\alpha_{k-1})}\,
\end{equation*}
which, for  $ \alpha_1 + \ldots +\alpha_{k-1} \neq\alpha_{k+1} +
\ldots +\alpha_{2k-1}$, implies that $\< \rho,\varphi\> < 0$. If
$\alpha_1 + \ldots + \alpha_{k-1} = 0$, then $\alpha_{k+1} +
\ldots + \alpha_{2k} = 1-\alpha_0-\alpha_k$ and
\[ G(a) = (1-a)(1-\alpha_0-\alpha_k)  \ . \]
Hence, $G(a) < 0$ iff $a > 1$ and $1> \alpha_0 +\alpha_k >0$.

\section{Conclusions}

We have constructed a new class of positive indecomposable maps
$\varphi : M_d \longrightarrow M_d$ which generalizes a Choi map
on $M_3$ \cite{Choi1}. Each such map is characterized by a cyclic
bistochastic $d \times d$ matrix $||a_{ij}||$ satisfying
conditions of Theorem~1. Now, any indecomposable map provides a
new tool for investigation of quantum entanglement: a PPT state
$\rho$ is entangled iff there exists an indecomposable map
$\varphi$ such that $(\oper_d \ot \varphi)\rho \ngeq 0$, i.e.
$(\oper_d \ot \varphi)\rho$ has at least one negative eigenvalue.
Recall that a characteristic feature of transposition $\tau$ is
that $\tau$ and $\tau_U$ defined by
\[   \tau_U(X) = U\, X^T\, U^\dag\ ,\]
for $U \in U(d)$, are equivalent, i.e. $(\oper_d \ot \tau)\rho$
and $(\oper_d \ot \tau_U)\rho$ have the same eigenvalues
\cite{Peres}. This property is no longer true for other positive
maps. In general $\varphi$ and $\varphi_U$:
\[  \varphi_U(X) = U\, \varphi(X)\, U^\dag\ ,  \]
 are not equivalent, that is,
even if $(\oper_d \ot \varphi)\rho\geq 0$ there may still exist $U
\in U(d)$ such that $(\oper_d \ot \varphi_U)\rho \ngeq 0$.

Therefore any indecomposable map $\varphi$ defined by
(\ref{GEN-b}) gives rise to the whole class of indecomposable maps
$\varphi_U$:
\begin{eqnarray}\label{}
\varphi_U(e_{ii}) &=& \sum_{j,k,l=1}^d  a_{ij}
U_{jk}\overline{U}_{jl}e_{kl} \ , \nonumber \\
 \varphi_U(e_{ij})
&=& - \frac{1}{d-1} \sum_{k,l=1}^d\, U_{ik}\overline{U}_{jl}\,
e_{kl} \ , \ \ \ i\neq j \ ,
\end{eqnarray}
with $U_{ik} = \mbox{Tr}(Ue_{ki})$. This construction leads to a
new family of indecomposable entanglement witnesses
\begin{equation*}\label{}
    \widehat{\varphi}_U = (\oper_d \ot \varphi_U)P^+ =
    \sum_{i,j=1}^d e_{ij} \ot\, U \varphi(e_{ij})\, U^\dag\ .
\end{equation*}
As a byproduct we showed that this family of indecomposable
entanglement witnesses  detect quantum entanglement within a large
class of PPT states proposed recently in \cite{PPT-nasza}.

\section*{Acknowledgement} This work was partially supported by the
Polish State Committee for Scientific Research Grant {\em
Informatyka i in\.zynieria kwantowa} No PBZ-Min-008/P03/03.

\section*{Appendix } \label{A}
\def\theequation{A.\arabic{equation}}
\setcounter{equation}{0}

Consider $d=3$. Any contraction in $d-1=2$ dimensions is
represented by $A = R_1 D R_2$, where $D$ is a diagonal matrix
with $D_{ii}=\lambda_i$ such that $|\lambda_i| \leq 1$ and $R_k$
are orthogonal $2\times 2$ matrices. Hence, $R_k$ may be
parameterized as follows:
\begin{equation}\label{}
    R_k = \left( \begin{array}{cr} \cos\phi_k & - \sin\phi_k \\
    \sin\phi_k & \cos\phi_k \end{array} \right)\ .
\end{equation}
Now, the corresponding bistochastic $3 \times 3$ matrix
$a=||a_{ij}||$ reads as follows:
\begin{equation}\label{}
    a = P_0 + P_1 + P_2\ ,
\end{equation}
where
\begin{equation}\label{}
    P_0 = \frac 13\, \left( \begin{array}{ccc} 1& 1& 1 \\ 1& 1& 1
    \\1& 1& 1 \end{array} \right)\ ,
\end{equation}
\begin{equation}\label{}
P_1 = \frac{\lambda_+}{12}\, \left( \begin{array}{ccc} 2\cos\phi_+
& -\cos\phi_+ - \sqrt{3}\sin\phi_+ & - \cos\phi_+ +
\sqrt{3}\sin\phi_+ \\
- \cos\phi_+ + \sqrt{3}\sin\phi_+ & 2\cos\phi_+ & -\cos\phi_+ -
\sqrt{3}\sin\phi_+  \\
-\cos\phi_+ - \sqrt{3}\sin\phi_+  & - \cos\phi_+ +
\sqrt{3}\sin\phi_+ & 2\cos\phi_+ \end{array} \right)\ ,
\end{equation}
and
\begin{equation}\label{}
P_2 = \frac{\lambda_-}{12}\, \left( \begin{array}{ccc} \cos\phi_-
+ \sqrt{3}\sin\phi_- & -  2\cos\phi_- & \cos\phi_- -
\sqrt{3}\sin\phi_- \\ -  2\cos\phi_- & \cos\phi_- -
\sqrt{3}\sin\phi_- & \cos\phi_- + \sqrt{3}\sin\phi_-
\\
\cos\phi_- - \sqrt{3}\sin\phi_- & \cos\phi_- + \sqrt{3}\sin\phi_-
& - 2\cos\phi_-  \end{array} \right)\ ,
\end{equation}
with $\phi_\pm = \phi_1 \pm \phi_2$ and $\lambda_\pm = \lambda_1
\pm \lambda_2$. Note that for $\lambda_1 = \lambda_2 = 1$ and
$\phi_1=0$, $\phi_2=-\pi/3$ one recovers (\ref{a-Choi}).

\end{document}